\documentclass[reprint, aps, prl, nofootinbib, superscriptaddress, fleqn]{revtex4-2}
\usepackage[utf8]{inputenc}
\usepackage{amsmath}
\usepackage{amssymb}
\usepackage{graphicx}
\usepackage[dvipsnames]{xcolor}
\usepackage{calc}
\usepackage{etoolbox}
\usepackage{booktabs}
\usepackage{tabularx}
\usepackage{multirow}
\usepackage[normalem]{ulem}
\usepackage{siunitx}
\usepackage{mleftright}
\usepackage{hyperref}
\usepackage[capitalise]{cleveref}

\makeatletter
\input{aas_macros.sty}
\let\jnl@style=\relax
\makeatother

\makeatletter
\AtBeginDocument{\immediate\write\@auxout{\string\citation{apsrev42Control}}}
\makeatother

\newlength{\affilindent}
\newcommand*{\affilscriptphantom}{\phantom{\normalfont\textsuperscript{99}}}
\newcommand*{\printaffiliation}[4]{%
    \ifnumequal{#1}{0}{}{%
        \setlength{\affilindent}{\widthof{\affilscriptphantom}}%
        \let\\=\newline%
        \par\noindent%
        \looseness=-100\hangindent=\affilindent%
        \affilscriptphantom\llap{\normalfont\textsuperscript{#1}}%
        \ignorespaces#3%
        \par%
    }%
}

\newcommand{\Msun}{ \,   M_{\odot}  {h}^{-1}   }

\renewcommand{\vec}[1]{\boldsymbol{#1}}

\newcommand{\six}[6]{\left(\begin{array}{ccc}
									{#1}& {#2}& {#3}\\
									{#4}& {#5}& {#6} \\
\end{array}\right)}

\newcommand{\tjo}[3]{\begin{pmatrix} {#1} & {#2} & {#3}\\ 0 & 0 & 0\end{pmatrix}}

\makeatletter

\DeclareSIUnit{\year}{yr}
\DeclareSIUnit{\Gyr}{\giga\yr}
\DeclareSIUnit{\pc}{pc}
\DeclareSIUnit{\kpc}{\kilo\pc}
\DeclareSIUnit{\Mpc}{\mega\pc}
\DeclareSIUnit{\Gpc}{\giga\pc}
\DeclareSIUnit{\hHubble}{\text{\ensuremath{h}}}
\DeclareSIUnit{\Msun}{\text{\ensuremath{M_\odot}}}
\newcommand*{\eqlabelleft}{(}
\newcommand*{\eqlabelright}{)}

\creflabelformat{equation}{\eqlabelleft#2#1#3\eqlabelright}

\begin{document}

\title{Can the CMB be Odd? Effect of a Parity-Violating Matter 4-Point Function on the Low-$\ell$ CMB Trispectrum}

\author{Zachary Slepian}
\affiliation{Department of Astronomy, University of Florida,\\211~Bryant Space Science Center, Gainesville, FL 32611, USA}

\author{Matthew Reinhard}
\affiliation{Department of Physics, University of Florida,\\2001 Museum Road, Gainesville, FL 32611, USA}

\author{Michael Bartlett}
\affiliation{Department of Physics and Astronomy, University of Oklahoma,\\440 West Brooks Street, Norman, OK 73019, USA}

\begin{abstract}
Here we explore from a theoretical perspective the sensitivity of the primary CMB anisotropy trispectrum to parity violation (PV) in large-scale structure (LSS). We focus on the Sachs-Wolfe term, which dominates at $\ell < 40$, after which the Doppler term takes over. We consider a model where the PV is only present out to some maximal scale $R$ of order a few hundred Mpc/$h$, consistent with what recent LSS 4PCF measurements seem to indicate. We find that the odd CMB trispectrum must be suppressed by at least one factor of $R/\chi_*$$\simeq$1\%, with $\chi_*$ the distance to last scattering, relative to the input matter 4PCF. Thus, a non-detection of PV in the CMB trispectrum is \textit{not necessarily inconsistent} with a genuine detection of PV in the LSS 4PCF.
\end{abstract}
						  
\maketitle

\section{Introduction}
\label{sec:intro}
Recent work has shown that the 4-Point Correlation Function (4PCF) is sensitive to parity violation (PV) in large-scale structure (LSS) \cite{cahn_short_pub}. The 4PCF measures excess spatial clustering over and above what a random distribution would have; see \textit{e.g.} \cite{slepian_encyc, phil_even_4pcf, hou_even_desi}.\footnote{\cite{jeong_clus_fossils_pub} presented a related statistic using pairs of power spectra, and recent work has explored using galaxy shapes \cite{coulton_pv_sims, kurita_2025}, Baryon Acoustic Oscillation (BAO) features \cite{bao_odd}, and compressed statistics \cite{jamieson_pop_pub, gao_2025, kurita_2025} for PV.} Evidence for PV was found in Sloan Digital Sky Survey Baryon Oscillation Spectroscopic Survey (SDSS BOSS; \cite{hou_parity_pub}, see also \cite{phil_parity})\footnote{Both of these relied heavily on a covariance matrix from \cite{hou2021analytic}.}, and more recently, in Dark Energy Spectroscopic Instrument (DESI) data \cite{desi_parity}. 

It is thus of interest to ask whether PV in the large-scale distribution of matter, if produced during inflation, and thus present at recombination, would imprint on the Cosmic Microwave Background (CMB). Notably, \cite{phil_cmb} examined the CMB temperature anisotropy trispectrum for PV and found no detection (see also \cite{phil_pol} for inclusion of polarization).\footnote{\cite{greco_lensing} explored the theoretical sensitivity of the CMB \textit{lensing} trispectrum to PV in LSS.} We might well ask---is this consistent with the evidence seen in 3D LSS? 

In this paper, we explore the impact of PV in the matter on the large-scale, low-$\ell$ CMB. We focus on this because in the low-$\ell$ limit, the CMB anisotropies are dominated by the Sachs-Wolfe (SW) effect \cite{sw_1967, sw_1997_white, baumann}, which is that photons leaving an over-dense region are redshifted and leaving an under-dense region are blueshifted. Thus on large scales the CMB temperature anisotropies trace the matter fluctuations, and the photon transfer function is simple and analytically tractable. In future work we will explore the higher-$\ell$ regime.

We write down a general expansion for the parity-violating 3D matter 4PCF, and from this compute the matter trispectrum by taking a Fourier Transform (FT). From this matter trispectrum we may then compute the trispectrum of the primary CMB anisotropies. 

We find that the CMB trispectrum can be expressed in terms of triple-spherical Bessel function (sBf) integrals that, for an odd-parity input matter 4PCF, will have leading behavior as $R/\chi_*$, where $R$ is the largest scale on which the PV 4PCF has support, and $\chi_*$ is the distance to last scattering. Since observationally, at present $R$$\sim$300 Mpc/$h$, while $\chi_* \approx 20,000$ Mpc/$h$, this shows that generically the primary CMB's sensitivity to PV is ``geometrically'' suppressed by a factor $\sim$100 relative to the input matter 4PCF. 

This paper is structured as follows. We first present the relationship between the CMB trispectrum and the primordial curvature perturbation trispectrum in the SW limit. We then relate the primordial trispectrum to the linear matter trispectrum, and the latter to the position-space matter 4PCF. By interchanging the order of integrations, we show that triple-sBf integrals enter, and we perform all of the angular integrations analytically. 

We next introduce an approximate matter transfer function that enables us to proceed with the radial integrals over the $k_i$, and show how to perform these analytically. We then examine the behavior of the results, showing it is, for an odd-parity input matter 4PCF, always suppressed by $R/\chi_*$. It is not always suppressed in this way for an even-parity input 4PCF. Finally, we provide some concluding discussion.

\section{CMB Trispectrum}
Consider the (connected) primary CMB trispectrum $\mathcal{T}$ in the Sachs-Wolfe limit:
\begin{align}
\label{eq:CMB_trispec}
&\mathcal{T}_{\ell_1 \ell_2 \ell_3 \ell_4}^{m_1 m_2 m_3 m_4} \equiv \left< a_{\ell_1 m_1} a_{\ell_2 m_2} a_{\ell_3 m_3} a_{\ell_4 m_4} \right>_{\rm c}\\
&= \int d^3\vec{k}_1 \cdots d^3\vec{k}_4\,
\frac{1}{5^4} j_{\ell_1}(k_1 \chi_*) \cdots j_{\ell_4}(k_4 \chi_*)\nonumber\\
&\qquad \times \delta_{\rm D}^{[3]}(\vec{k}_1 + \cdots + \vec{k}_4)
Y_{\ell_1 m_1}^*(\hat{k}_1)\cdots Y_{\ell_4 m_4}^*(\hat{k}_4)\nonumber\\
&\qquad \times \mathcal{T}_{\rm pri}(\vec{k}_1, \vec{k}_2, \vec{k}_3, \vec{k}_4).\nonumber
\end{align}
Conventionally one imposes $\ell_1 \leq \ell_2$ etc. to avoid redundancy of information (\textit{e.g.} \cite{hu_cmb_trispectrum, okamoto_hu}). Subscript $\rm c$ is for ``connected.'' $\chi_*$ is the distance to last scattering, and the second line is the set of four SW transfer functions. $\delta_{\rm D}^{[3]}$ is a 3D Dirac delta function, expressing 3D translation invariance of the underlying matter density fluctuations. Each temperature anisotropy is projected onto a harmonic, resulting in the $Y_{\ell_i m_i}^*$ in the third line. The last line, $\mathcal{T}_{\rm pri}$, is the primordial matter trispectrum.

The $a_{\ell_i m_i}$ are the harmonic coefficients of the CMB temperature anisotropies, defined via
\begin{align}
    a_{\ell_i m_i} &\equiv 4\pi i^{\ell_i} \int d\hat{n}_i\;\delta T(\hat{n_i}) Y_{\ell_i m_i}^*(\hat{n}_i)\nonumber\\
    &= \int \frac{d^3\vec{k}}{(2\pi)^3}\;\Theta_{\ell}(k)\mathcal{R}_{\rm pri}(\vec{k}) Y_{\ell_i m_i}^*(\hat{k}) 
    \label{eq:alm_def}
\end{align}
with $\delta T (\hat{n}_i) \equiv T(\hat{n}_i)/\bar{T} - 1$ the relative temperature fluctuation in the direction $\hat{n}_i$. $T$ is the temperature and bar denotes mean. In the second line, we have expressed the $a_{\ell m}$ in terms of the photon transfer function, $\Theta_{\ell}$, and the primordial curvature perturbation, $\mathcal{R}_{\rm pri}$ \cite{baumann}. We note that this relation is general and does not yet assume the SW limit, $\Theta_{\ell}(k) \to (1/5) j_{\ell}(k \chi_*)$, which we have taken in  Eq. \ref{eq:CMB_trispec}.

Now, the trispectrum of the primordial curvature perturbations $\mathcal{R}_{\rm pri}$ is
\begin{align}
    \label{eq:Tpri_Tlin}
    &\mathcal{T}_{\rm pri}(\vec{k}_1, \ldots, \vec{k}_4) 
    \equiv \left<\mathcal{R}_{\rm pri}(\vec{k}_1) \cdots  \mathcal{R}_{\rm pri}(\vec{k}_4) \right>
    \propto \\
    &\mathcal{T}_{\rm lin}(\vec{k}_1, \ldots, \vec{k}_4) T_{\rm m}^{-1}(k_1) \cdots T_{\rm m}^{-1}(k_4) ,\nonumber
\end{align}
with $\mathcal{T}_{\rm lin}$ the linear matter trispectrum and $T_{\rm m}$ the matter transfer function. The proportionality rather than equality is because we neglect four factors of the linear growth rate, as they will not affect the scale-dependence in this calculation.

Now our linear trispectrum is, by definition, related to the linear matter 4PCF by a Fourier Transform (FT), and so we have
\begin{align}
    &\mathcal{T}_{\rm lin}(\vec{k}_1, \ldots, \vec{k}_4) (2\pi)^3 \delta_{\rm D}^{[3]}(\vec{k}_1 +\cdots + \vec{k}_4) \nonumber\\
    &=\int d^3\vec{x}\, d^3\vec{r}_1 \cdots d^3\vec{r}_3\; e^{- i \vec{k}_4 \cdot \vec{x}}\nonumber\\
    &\qquad \times e^{- i \vec{k}_1 \cdot (\vec{x} + \vec{r}_1)}
    e^{- i \vec{k}_2 \cdot (\vec{x} + \vec{r}_2)}
    e^{- i \vec{k}_3 \cdot (\vec{x} + \vec{r}_3)}\nonumber\\
    &\qquad \times \left<\delta_{\rm lin}(\vec{x}) \delta_{\rm lin}(\vec{x} + \vec{r}_1) \cdots \delta_{\rm lin}(\vec{x} + \vec{r}_3) \right>_{\rm c}.
    \label{eq:T_lin}
\end{align}
By convention, the delta function on the left-hand side is written explicitly and not included in the definition of the trispectrum itself, but always follows along with it.

We note that the brackets in the last line represent an expectation value over realizations of the Universe. While it is conventionally taken to be equivalent to averaging over $d^3\vec{x}/V$, with $V$ the integration volume, by ergodicity, conceptually it is distinct and so we indicate it explicitly here.

Now, the expectation value in the last line is simply the connected 4PCF, which we denote $\zeta(\vec{r}_1, \vec{r}_2, \vec{r}_3)$ (suppressing subscript c for brevity). Expanding it into the isotropic basis \cite{iso_basis_pub, iso_gen_pub}, where the ordering of magnitudes $r_1 < r_2 < r_3$ is necessary to avoid over-completeness of the basis, we have
\begin{align}
    \label{eq:zeta}
    &\zeta(\vec{r}_1, \vec{r}_2, \vec{r}_3) = \\
    &\qquad \sum_{\ell_1',\ell_2', \ell_3'} \mathcal{P}_{\ell_1' \ell_2' \ell_3'}(\hat{r}_1, \hat{r}_2, \hat{r}_3) \zeta_{\ell_1' \ell_2' \ell_3'}(r_1, r_2, r_3). \nonumber
\end{align}
Now, inserting this into Eq. \eqref{eq:T_lin}, evaluating the $d^3\vec{x}$ integral to yield a delta function, inserting Eq. \eqref{eq:T_lin} into Eq. \eqref{eq:Tpri_Tlin}, and then what results into Eq. \eqref{eq:CMB_trispec}, we find
\begin{align}
    \label{eq:trispec_with_gaunts}
    &\mathcal{T}_{\ell_1 \ell_2 \ell_3 \ell_4}^{m_1 m_2 m_3 m_4}
    =\sum_{L_i}\int r^2 dr \nonumber\\
    &\times \iiint_{r_1<r_2<r_3}  r_1^2\, dr_1
    \,r_2^2\, dr_2\,r_1^3\, dr_3\,\zeta_{\ell_1' \ell_2' \ell_3'}(r_1, r_2, r_3)\nonumber\\
    &\times I_{\ell_1 \ell_1' L_1}(\chi_*, r_1, r)
    I_{\ell_2 \ell_2' L_2}(\chi_*, r_2, r)
    I_{\ell_3 \ell_3' L_3}(\chi_*, r_3, r)\nonumber\\
    &\times \mathcal{I}_{\ell_4 L_4}(\chi_*, r)\nonumber\\
    &\times \mathcal{G}_{\ell_1 \ell_1' L_1}^{-m_1 m_1' M_1}
    \mathcal{G}_{\ell_2 \ell_2' L_2}^{-m_2 m_2' M_2}
    \mathcal{G}_{\ell_3 \ell_3' L_3}^{-m_3 m_3' M_3}  (-1)^{m_1 + m_2 + m_3} \nonumber\\
    &\times \delta^{\rm K}_{\ell_4 L_4}\mathcal{H}_{L_1 L_2  L_3 L_4}^{M_1 M_2  M_3 M_4}.
\end{align}
We have defined
\begin{align}
    &I_{\ell_i \ell'_i L_i}(\chi_*, r_i, r) \equiv \nonumber\\
    &\qquad \int k_i^2 dk_i\; j_{\ell_i}(k_i \chi_*) j_{\ell'_i}(k_i r_i) j_{L_i}(k_i r) T_{\rm m}^{-1}(k_i),\nonumber\\
    &\mathcal{I}_{\ell_i L_i}(\chi_*, r)  \equiv \int k_i^2 dk_i\; j_{\ell_i}(k_i \chi_*) j_{L_i}(k_i r) T_{\rm m}^{-1}.
\end{align}
The $\mathcal{H}$ is from the expansion of the four-argument delta function (\textit{e.g.} \cite{pt_decoupling}), while the Gaunts are from expanding the isotropic basis function in the trispectrum (resulting from that in the 4PCF), and integrating against the harmonic from the delta function and from the projection for each argument $\hat{k}_1, \hat{k}_2, \hat{k}_3$. 

We note that for $\hat{k}_4$, we have no harmonic from the trispectrum as there is no dependence in the 4PCF on the position vector conjugate to $\vec{k}_4$. Put another way, the trispectrum's dependence on its internal angles can be written as a function of $\hat{k}_1, \hat{k}_2, \hat{k}_3$ and $\hat{k}_4$ is not explicitly required (which makes sense given the delta function). 

\section{Approximate Matter Transfer Function}
To proceed further analytically, we need a form for the matter transfer function that can be easily inverted. We will use an approximate model wherein
\begin{align}
    T_{\rm m}^{-1}(k) \approx 1 + (k / k_{\rm eq})^2,
    \label{eq:Tm_approx}
\end{align}
with $k_{\rm eq} \approx 0.01\,h/$ Mpc the wave-number associated with the sound horizon at matter-radiation equality (this form is motivated by \cite{eisenstein_hu_1998}). This transfer function misses the BAO wiggles, which are a few-percent feature in the matter transfer function, but it is accurate as regards the overall shape of the transfer function, and captures the smooth quadratically-falling knee around $k_{\rm  eq}$.

\section{Delta Function Expansion}
By writing the delta function as an FT of unity, and then applying the plane wave expansion four times and integrating over directions (\textit{e.g.} as done in \cite{se_3pt_alg} for the three-argument delta function, and similar to the treatment of the four-argument delta function in \cite{iso_gen_pub}), we have
\begin{align}
\label{eq:delta_func}
    &(2\pi)^3 \delta_{\rm D}^{[3]}(\vec{k}_1 + \cdots + \vec{k}_4) = \int d^3 \vec{r}\; e^{i\vec{k}_1 \cdot \vec{r}}
    \times \cdots \times e^{i\vec{k}_4 \cdot \vec{r}}\nonumber\\
    &=(4\pi)^4\sum_{L_i M_i} i^{L_1 + \cdots + L_4}\nonumber\\
    &\times \int d\Omega_r\; Y^*_{L_1 M_1}(\hat{r}) \times \cdots \times Y^*_{L_4 M_4}(\hat{r})\nonumber\\
    &\times\int r^2\, dr\;j_{L_1}(k_1 r) \times \cdots \times j_{L_4}(k_4 r) \nonumber\\
    &\times Y_{L_1 M_1}(\hat{k}_1)\times \cdots \times  Y_{L_4 M_4}(\hat{k}_4)
\end{align}
The integral over four harmonics in $\hat{r}$ is simply \cite{pt_decoupling}
\begin{align}
    &\mathcal{H}_{L_1 L_2 L_3 L_4}^{M_1 M_2 M_3 M_4} \equiv \\
    &\qquad \sum_{L_{12}, M_{12}} \mathcal{G}_{L_1 L_2 L_{12}}^{M_1 M_2 -M_{12}}
    \mathcal{G}_{L_{12} L_3 L_4}^{M_{12} M_3 M_4}.\nonumber
\end{align}
We will use this by splitting up the $r$ integral in the fourth line above to enable doing the $k_i$ integrals in our trispectrum expression first. Denoting the normalization and phase as
\begin{align}
    \mathcal{D}_{L_1 L_2 L_3 L_4} \equiv (4 \pi)^4 i^{L_1 + L_2 + L_3 + L_4}
\end{align}
enables us to write the Delta function compactly. We also use the definition $\mathcal{R}_{L_1 L_2 L_3 L_4}$ to denote the four-sBf integral in Eq. (\ref{eq:delta_func}), and finally, we define for later use the product of three sBfs:
\begin{align}
\label{eq:calJ}
    &\mathcal{J}_{L_1 L_2 L_3}(k_1, k_2, k_3; r) \equiv\\
    &\qquad \qquad j_{L_1}(k_1 r) j_{L_2}(k_2 r) j_{L_3}(k_3r).\nonumber
\end{align}

\section{Innermost Integral}
Now, we may evaluate the integral over $k_4$. We have
\begin{align}
\label{eq:k4_decomp}
    &I_{4,\ell_4}(\chi_*, r) \equiv \frac{1}{5} \delta^{\rm K}_{\ell_4 L_4}   \nonumber\\
    &\times \int \frac{k_4^2 dk_4}{2 \pi^2}\, j_{\ell_4}(k_4 \chi_*) j_{L_4} (k_4 r)\, [ 1 + (k_4/k_{\rm eq})^2]\nonumber\\
    &\equiv I_{4,1} + k_{\rm eq}^{-2} I_{4,2},
\end{align}
where $I_{4,1}$ represents the contribution from unity in the square brackets and $I_{4,2}$ that from the quadratic term in the square brackets. For brevity we will suppress the arguments and the $\ell_4$-dependence during our intermediate steps.

Straightforwardly we have
\begin{align}
    I_{4,1} = \frac{1}{20 \pi} \frac{1}{r \chi_*} \delta_{\rm D}^{[1]}(r - \chi_*)
    \label{eq:I41}
\end{align}
from the term proportional to unity in the square brackets in Eq. (\ref{eq:k4_decomp}); this is simply the sBf orthogonality relation. Had we chosen to write $1/r^2$ in the pre-factor above, we have verified that our final results will be unchanged. 


The term proportional to $k_4^2$ in the square brackets may be evaluated using a method presented in \cite{assassi} (see also \cite{meigs}), wherein one applies the sBf differential equation as a parametric differentiation operator, $\hat{D}_{\ell_4, \chi_*}$, to bring in two powers of $k_4$, converting a result with weight $k_4^2$ to one with weight $k_4^4$. We have
\begin{align}
    I_{4,2} =\hat{D}_{\ell_4, \chi_*} I_{4,1}
\end{align}
where
\begin{align}
    \hat{D}_{\ell_4, \chi_*} \equiv -\left[\frac{\partial^2}{\partial \chi_*^2} + \frac{2}{\chi_*}\frac{\partial}{\partial \chi_*} - \frac{\ell_4 (\ell_4 +1)}{\chi_*^2} \right].
    \label{eq:assassi}
\end{align}
Applying this operator to Eq. \ref{eq:I41}, we find
\begin{align}
&\hat{D}_{\ell_4, \chi_*} \left[  \frac{1}{20\pi r \chi_*} \delta_{\rm D}^{[1]}(\chi_* - r)\right] = -\frac{1}{20 \pi r }\\
    &\times \bigg\{\chi_*^{-1} \delta_{\rm  D}''(\chi_* - r) - \ell_4 (\ell_4 + 1) \chi_*^{-3}\delta_{\rm D}(\chi_* - r) \bigg\},\nonumber
\end{align}
where prime denotes $\partial/\partial \chi_*$. We may see easily that the dimensions of each term are the same above. We note that the derivative of the delta function is defined by integration by parts \cite{Barton, amaku_1, amaku_2}. For test functions with appropriate conditions, it simply gives the test function's derivative evaluated at that point.\footnote{See also \url{https://mathworld.wolfram.com/DeltaFunction.html}.}

Applying this result within $\mathcal{R}_{L_1 L_2 L_3 L_4}$ (the four-sBf integral in Eq. \ref{eq:delta_func}) we have
\begin{align}
\label{eq:I42imp}
    &I_{4,2} \implies -\frac{1}{20 \pi }\left\{\chi_*^{-2}\ell_4 (\ell_4 + 1)+ \chi_*^{-1}\frac{\partial^2}{\partial \chi_*^2}\chi_* \right\}\nonumber\\
    &\qquad \qquad \qquad \mathcal{J}_{L_1 L_2 L_3}(k_1, k_2, k_3; \chi_*),
\end{align}
with $\mathcal{J}_{L_1 L_2 L_3}$ defined in Eq. (\ref{eq:calJ}).

Applying $I_{4,1}$ to $\mathcal{R}_{L_1 L_2 L_3}$ simply yields 
\begin{align}
\label{eq:I41imp}
    I_{4,1} \implies \frac{1}{20\pi} \mathcal{J}_{L_1 L_2 L_3}(k_1, k_2, k_3; \chi_*).
\end{align}
Combining Eqs. (\ref{eq:I42imp}) and (\ref{eq:I41imp}) as prescribed by Eq. (\ref{eq:k4_decomp}), we find that
\begin{align}
\label{eq:I4imp}
    & I_4 \implies \frac{1}{20\pi} [1 - A_{\ell_4} - B \frac{\partial^2}{\partial \chi_*^2} \chi_* ] \nonumber\\
    &\qquad\qquad\qquad \mathcal{J}_{L_1 L_2 L_3}(k_1, k_2, k_3; \chi_*) \nonumber\\
    &\equiv \hat{F}_{\ell_4,\chi_*} \mathcal{J}_{L_1 L_2 L_3}(k_1, k_2, k_3; \chi_*),
\end{align}
where the last line defines $\hat{F}$. We have also defined constants
\begin{align}
    &A_{\ell_4} \equiv \epsilon^2 \ell_4 (\ell_4 + 1),\;\; \epsilon \equiv r_{\rm eq}/\chi_* \ll 1, \nonumber\\
    &B \equiv \epsilon\, r_{\rm eq}, 
\end{align}
where we note that $B$ has dimensions of length, and thus the term in Eq. (\ref{eq:I4imp}) that includes $B$ remains dimensionless.

With the innermost, $k_4$, integral performed, we now have
\begin{align}
    \label{eq:trispec_with_gaunts_and_F_hat}
    &\mathcal{T}_{\ell_1 \ell_2 \ell_3 \ell_4}^{m_1 m_2 m_3 m_4}
    =  \hat{F}_{\ell_4, \chi_*}  \sum_{L_1 L_2 L_3}\nonumber\\
    &\iiint_{r_1<r_2<r_3}  r_1^2\, dr_1
    \,r_2^2\, dr_2\,r_1^3\, dr_3\nonumber\\
    &I_{\ell_1 \ell_1' L_1}(\chi_*, r_1, \chi_*)I_{\ell_2 \ell_2' L_2}(\chi_*, r_2, \chi_*)
    I_{\ell_3 \ell_3' L_3}(\chi_*, r_3, \chi_*)
    \nonumber\\
    &\times \mathcal{G}_{\ell_1 \ell_1' L_1}^{-m_1 m_1' M_1}
    \mathcal{G}_{\ell_2 \ell_2' L_2}^{-m_2 m_2' M_2}
    \mathcal{G}_{\ell_3 \ell_3' L_3}^{-m_3 m_3' M_3}  (-1)^{m_1 + m_2 + m_3} \nonumber\\
    &\times \mathcal{H}_{L_1 L_2 L_3 \ell_4}^{M_1 M_2 M_3 -m_4}\mathcal{D}_{L_1 L_2 L_3 \ell_4}.
\end{align}

\section{Final Result}
Now, each of the $I$ can be written as $[1 + k_{\rm eq}^{-2}\hat{D}_{\ell_i, r_i}]$ acting on a triple-sBf integral with $k_i^2$ weight. This latter integral was performed in \cite{Mehrem1991} (see also \cite{jackson_maximon_1972, mehrem_2011, Grant_1993, Fabrikant2013}) and we denote it $\mathcal{M}_{\ell_i \ell_i' L_i}$.\footnote{\cite{Grant_1993} gives an explicit formula for this integral in simple powers of the three free arguments, but we find the formula of \cite{Mehrem1991} easier to manipulate for the present work.} Our final result is thus
\begin{align}
    &\mathcal{T}_{\ell_1 \ell_2 \ell_3 \ell_4}^{m_1 m_2 m_3 m_4} = \hat{F}_{\ell_4, \chi_*}  \iiint_{r_1 < r_2 < r_3} \nonumber\\ 
    &[1 + k_{\rm eq}^{-2} \hat{D}_{\ell_1, r_1}] 
    [1 + k_{\rm eq}^{-2} \hat{D}_{\ell_2, r_2}]
    [1 + k_{\rm eq}^{-2} \hat{D}_{\ell_3, r_3}]\nonumber\\
    &\times \sum_{L_1 L_2 L_3} \mathcal{M}_{\ell_1 \ell_1' L_1}
    \cdots
    \mathcal{M}_{\ell_3 \ell_3' L_3}\nonumber\\
    &\times \mathcal{G}_{\ell_1 \ell_1' L_1}^{-m_1 m_1' M_1}
    \mathcal{G}_{\ell_2 \ell_2' L_2}^{-m_2 m_2' M_2}
    \mathcal{G}_{\ell_3 \ell_3' L_3}^{-m_3 m_3' M_3}  (-1)^{m_1 + m_2 + m_3} \nonumber\\
    &\times \mathcal{H}_{L_1 L_2 L_3 \ell_4}^{M_1 M_2 M_3 -m_4} \mathcal{D}_{L_1 L_2 L_3 \ell_4}.
\end{align}
$\hat{D}$ is defined in Eq. \ref{eq:assassi}.

\section{Behavior of the $\mathcal{M}$ Integrals}
We know that the sum of the $\ell_i'$ is odd. We also know that each index ($i = 1,2,3$) has an even sum of the $\ell_i, \ell_i', L_i$ tied to it, from the Gaunts. We then know that $\ell_i$ and $L_i$ must have opposite parity. This results in a suppression in each $\mathcal{M}$ integral result. We show this explicitly below.

We have, evaluating \cite{Mehrem1991, mehrem_2011}'s result in the limit of two free arguments' being equal, that
\begin{align}
    &\mathcal{M}_{\ell_i \ell_i' L_i} = \tjo{\ell_i}{\ell_i'}{L_i}^{-1}
    \frac{\pi \beta(\Delta)}{4 \chi_*^2 r_i} i^{\ell_i + \ell_i' + L_i} \nonumber\\
    &\times \sqrt{2L_1 + 1}
    (-1)^{\ell_i + L_i}\sum_{L = 0}^{L_i} {2 L_i \choose 2L}^{1/2} \left(\frac{r_i}{\chi_*} \right)^L \nonumber\\
    &\times \sum_{\ell} (2\ell + 1)\tjo{\ell_i}{L_i - L}{\ell} \tjo{\ell_i'}{L}{\ell} \nonumber\\
    &\times \six{\ell_i}{\ell_i'}{L_i}{L}{L_i - L}{\ell} \mathcal{L}_{\ell}(\Delta),\;\;\;\Delta \equiv \frac{1}{2}\frac{r_i}{\chi_*}.
\end{align}
Let us now examine the behavior of this result with $L$, an intermediate angula momentum that is summed over. We see that if $L = 1$ or higher, then our result is suppressed by some power of $r_i/\chi_*$. Thus, the lowest-order behavior of our result in $r_i/\chi_*$ will always be given by $L = 0$. But the 3-$j$ symbol in $\ell_i', L, \ell$ means that $\ell_i' + L + \ell$ is even. We now recall that by construction, $\ell_1' + \ell_2' + \ell_3'$ is odd, as we input a parity-odd 4PCF originally. Thus, either one or three of the $\ell_i'$ must be odd itself. 

We now examine what happens when $\ell_i'$ is odd. Since $\ell_i' + L + \ell$ is even, and we have set $L = 0$ to get the lowest-order behavior in $r_i/\chi_*$, we see that $\ell$ must be odd. But for odd $\ell$, the lowest order behavior of the Legendre polynomial will be as $\Delta$, and $\Delta \propto r_i/\chi_*$. Thus, for at least one $\ell_i'$, we have shown the CMB trispectrum is suppressed by at least $r_i/\chi_*$; at least, this is true for the terms proportional to unity in $\hat{F}$ and in $[1 + k_{\rm eq}^{-2}\hat{D}]$. 

\subsection{Term with $\hat{D}$}
We now ask what happens from the term involving $\hat{D}$. Dimensionally, all of these terms act like lowering the power of $r_i$ by two, and in fact this behavior is exactly what happens if we write out the integral $\mathcal{M}$ in a finite power series by expanding the Legendre in powers of $\Delta$ via Wolfram Eq. 32.\footnote{\url{https://mathworld.wolfram.com/LegendrePolynomial.html}} 

Thus we might worry that the terms with $\hat{D}$ eliminate the suppression we have found. However, we notice that all of these terms will also be proportional to $k_{\rm eq}^{-2} = r_{\rm eq}^2$, whereas the terms in the square brackets proportional to unity are not. Thus, this factor relative to $\chi_*^2$ restores the suppression, since $r_{\rm eq}$ will be of the same order as the $r_i$ we consider. 

\subsection{Effect of $\hat{F}$}
We now consider the effect of $\hat{F}$. The part of $\hat{F}$ that involves $1-A$ will have no additional effect on the power of $r_i/\chi_*$ we have. The part of it that involves a derivative operator is a derivative with respect to $\chi_*$; this will only increase the powers of $\chi_*$ in the denominator, hence increasing the suppression.  

\subsection{Even-Parity Input 4PCF}
We now briefly show that there is not necessarily a suppression if the input matter 4PCF is even-parity, \textit{i.e.} that the odd CMB trispectrum will be suppressed relative to some cases of the even one. 

Now, for an even input matter 4PCF, either one or three of the $\ell_i'$ must be even.  We ask what occurs when $\ell_i'$ is even, focusing for simplicity on what results from the non-derivative parts of the $\hat{F}$ and $[1 + \hat{D}]$ terms (\textit{i.e.} just the $\mathcal{M}$ integral). Here, the leading term within $\mathcal{M}$ is still that where $L = 0$, but now $\ell$ must be even. Thus the leading power of $\Delta$ is $\Delta^0 = 1$, so, for each even $\ell_i'$, there is no suppression. This shows that one can have an even-parity input matter 4PCF with all three $\ell_i'
$ even with no ``geometric'' $R/\chi_*$ suppression. We note that an overall-even input 4PCF with only one of its three $\ell_i'$ even will be suppressed as $(R/\chi_*)^2$.

\subsection{Divergence of the $(k_i/k_{\rm eq})^2$ Integral}
We now briefly treat a more subtle issue that arises in our calculation. From analysis of the UV asymptotics of the sBfs, we see that the parts of the $k_i$, $i=1,2,3$ integrals that involve $(k_i/k_{\rm eq})^2$ are divergent, since the integrand will scale overall as $k_i$ in the UV. Yet, $\mathcal{M}$ is everywhere finite, but we are using parametric differentiation of $\mathcal{M}$ to obtain the $(k/k_{\rm eq})^2$ result.

So how does the divergence arise? We note that $\mathcal{M}$ has a discontinuity when one transitions from a non-degenerate to a degenerate triangle; $\beta = 1$ if the three free arguments of $\mathcal{M}$ are able to form a closed triangle, but $\beta = 1/2$ if this triangle is degenerate (zero area), and $\beta = 0$ otherwise. Taking the derivative of this ``jump'' discontinuity yields a divergent result. 

Notably, this discontinuity is going to be proportional to a Heaviside function, so we can see that acting with $\hat{D}$ on it will produce a delta function and the first derivative of a delta function; these are integrable singularities when we then perform the final integrals over the $r_i$. Thus, while at an intermediate step the $(k_i/k_{\rm eq})^2$ piece produces a divergence, its final contribution to the CMB trispectrum is finite.

\section{Concluding Discussion}
It is worth noting that the SW effect dominates the CMB power spectrum by itself only at $\ell < 40$; after that the Doppler term from baryon velocities takes over, though the SW term is still important and at times exceeds the Doppler term; this implies the SW transfer function is not sufficiently accurate alone to use at $\ell_i > 40$. Thus, our present calculation should be taken to apply only to the low-$\ell$ regime. 

However, this is the regime most relevant for studies of PV with the CMB primary anisotropies, as at higher $\ell$ the sky is effectively flat (since these $\ell$ correspond to physical fluctuations that are much smaller than the CMB shell's radius of curvature). Some discussion of this point is in Appendix C of \cite{greco_lensing}. They analyze the maximum magnitude of the scalar triple product (to which all the parity-odd basis functions are proportional) that can stem from four matter density points on the CMB's spherical shell at last scattering. They provide a purely geometric proof that this is of order 1\%; indeed, it is $\sim$$R/\chi_*$. 

In the present paper we have taken this intuition, about a parity-odd configuration of matter perturbations that could source a CMB trispectrum, and quantitatively connected it to the actual observable, the CMB trispectrum.   

In closing, we note that a fully analytical approach may be challenging at higher $\ell$ as the photon transfer function assumes a less simple form (\textit{e.g.} \cite{baumann}). At very high $\ell \gtrsim 3,000$, Silk (photon diffusion) damping renders the primary CMB smooth and so one would not expect to be able to use it for parity at such $\ell$. 

In future work we will seek an analytical approach to the regime $40 \lesssim \ell \lesssim 3,000$, where Silk damping has not yet taken over but the approximation adopted in this work, that the photon transfer function is purely from the SW effect, is not sufficient. We note that in this regime the Limber approximation can be used on the sBfs entering the calculation, and we speculate that this will require all sBfs to be in phase (and the phase of each goes as $\ell_i$) and thus force an even-parity CMB trispectrum.

\clearpage
\section*{Appendix}
\subsection{Reduced Trispectrum}
Here, we present an alternative approach to the central result of this work. We compute the reduced trispectrum \cite{hu_cmb_trispectrum, slepian_constrained_realization}, which is the sum of the product of $a_{\ell_i m_i}$ presented in the main text against two 3-$j$ symbols. In the reduced trispectrum the $m_i$ vanish from the problem, as isotropy demands, and we have a function simply of total angular momenta $\ell_i$, as 
\begin{align}
&Q_{\ell_1 \ell_2}^{\ell_3 \ell_4}(\ell_{12}) \equiv \sum_{m_i, m_{12}}
\six{\ell_1}{\ell_2}{\ell_{12}}{m_1}{m_2}{-m_{12}}\\
&\times \six{\ell_{12}}{\ell_3}{\ell_4}{m_{12}}{m_3}{m_4}\nonumber
\left<a_{\ell_1 m_1} a_{\ell_2 m_2} 
a_{\ell_3 m_3} 
a_{\ell_4 m_4} \right>.     
\end{align}
$\ell_{12}$ is necessary to fix how $\ell_1, \ell_2$ are coupled to $\ell_3, \ell_4$, e.g. \cite{hu_cmb_trispectrum} Sec. II.C.

Performing this sum on the right-hand side of Eq. (\ref{eq:CMB_trispec}) converts the spherical harmonics into a 4-argument isotropic basis function in the $\hat{k}_i$ \cite{iso_basis_pub, iso_gen_pub}. The delta function may be written as an FT of unity. We have
\begin{align}
    &Q_{\ell_1 \ell_2}^{\ell_3 \ell_4}(\ell_{12})
    = \int d^3\vec{k}_1 \cdots d^3\vec{k}_4 \; \nonumber\\ 
    &\times \frac{1}{5^4} j_{\ell_1}(k_1 \chi_*) \cdots j_{\ell_4}(k_4 \chi_*)\\
    &\times \int d^3 \vec{r}\, e^{i \vec{r} \cdot (\vec{k}_1 + \cdots + \vec{k}_4)}\, \mathcal{P}_{\Lambda}(\hat{K}) \sum_{\Lambda'} \mathcal{P}_{\Lambda'}(\hat{K}) \mathcal{T}_{\Lambda'}(K)\nonumber
\end{align}
where $\Lambda \equiv  L_1, L_2, L_{12}, L_3, L_4$, $\Lambda' \equiv \ell_1', \ell_2', \ell_{12}', \ell_3', \ell_4'$, and $\hat{K}$ and $K$ are the combinations respectively of all unit vectors in $\hat{k}_i$, and all $k_i$ magnitudes, but for $i=1,2,3$. $\mathcal{T}_{\Lambda'}$ is a wave-vector-magnitude-dependent coefficient capturing the expansion of the primordial trispectrum into the 3-argument isotropic basis functions. 

Now, using the relationship between 3 and 4-argument basis functions when one angular momentum of the 4-argument basis function is zero \cite{iso_basis_pub}, we promote the 3-argument basis function into a 4-argument one. We may then rewrite the resulting product of two 4-argument isotropic basis functions into a sum over single basis functions via the generalized Gaunt integral \cite{iso_basis_pub}. We find
\begin{align}
     &Q_{\ell_1 \ell_2}^{\ell_3 \ell_4}(\ell_{12}) = \sum_{\Lambda', \Lambda''} \int r^2 dr\; k_1^2 \,dk_1\cdots k_4^2 \, dk_4\\
     &\times j_{\ell_1''}(k_1 r)\cdots j_{\ell_4''}(k_4 r) \frac{1}{5^4} j_{\ell_1}(k_1 \chi_*) \cdots j_{\ell_4}(k_4 \chi_*)\nonumber\\
     &\qquad\times \mathcal{G}_{\Lambda, \Lambda', \Lambda''}\mathcal{T}_{\Lambda'}(K).\nonumber
\end{align}

\subsection{Using Asymptotics}
Here we present another way to see that the odd CMB trispectrum is suppressed relative to the connected even-parity one. As in the main text, we will find that the key to this conclusion is that $r_i \ll \chi_*$. 

Returning to Eq. (\ref{eq:trispec_with_gaunts_and_F_hat}), we can change variables in the integrals $I$ to $u_i = k_i r_i$. Then the two arguments of each integral that involve $\chi_*$ will become $u_i (\chi_*/r_i)$, and we may look at asymptotics. We  have for each integral $(i=1,2,3)$:
\begin{align}
    I_{\ell_i \ell_i' L_i} \to j_{\ell_i}(u_i (\chi_*/r_i)) j_{\ell_i'}(u_i) j_{L_i}(u_i  (\chi_*/r_i)).
\end{align}
We are interested in the large-argument asymptotics because $\chi_*/r_i  \gg 1$; we have
\begin{align}
    j_{\ell}(x) \to \frac{\sin (x - \ell \pi/2)}{x},\;\;\;x \to \infty.
\end{align}
We see that for the two sBfs with large arguments to be in phase, $L_i$ and $\ell_i$ must have the same parity. But this implies that $\ell_i'$ must be even, since we have a 3-$j$ symbol with zero lower row connecting $\ell_i, \ell_i', L_i$ (from the Gaunt integral in Eq. \ref{eq:trispec_with_gaunts}). If $\ell_i'$ is even for $i=1,2,3$, as this condition demands, then the 4PCF must be even. We thus see that to get in-phase behavior of the sBfs in the asymptotic limit, which is achieved quickly since $\chi_*/r_i \gg 1$ (see discussion below), is only possible for an even input 4PCF.

In contrast, if the 4PCF is odd, at least one of the $\ell_i'$ must be odd, and so for at least one of the sBf integrals, the 3-$j$ symbol from the Gaunt will demand that $\ell_i$ and $L_i$ have opposite parity and the asymptotic behavior of the sBfs will be out of phase, substantially suppressing the odd CMB trispectrum relative to the even one. 


We may ask how quickly this asymptotic is achieved; in other words, when may we start to apply the argument above to the integrand? For $u_i (\chi_*/r_i)\gtrsim 10$, which is a reasonable starting point to use the asymptotic, we must have $u_i > 10(r_i/\chi_*) \approx 0.1$. Since $u_i = k_i r_i$, if we assume that the range of $r_i$ we are interested in is $r_i > 10$ Mpc/$h$ (on smaller scales photon diffusion damping (Silk damping) will wash out any fluctuations in the matter left over from inflation), then we must have $k_i > 0.01$ Mpc/$h$. 

In short, our asymptotic for the two sBfs is applicable for wave-numbers greater than this $0.01$ Mpc/$h$. A primordial 4PCF whose dominant support is on scales less than a few hundred Mpc will imply the integrand here's dominant contribution indeed is in the range where our asymptotic is good. 

Observationally, the finding of \cite{hou_parity_pub, desi_parity, phil_parity} is that the \textit{linear} 4PCF is supported out to these scales, with no measurement having been made beyond a few hundred Mpc/$h$. Linear evolution will not greatly alter the range of support. Thus observationally, all we know is that the primordial 4PCF is supported at least up to a few hundred Mpc/$h$, with no measurement made beyond. 

Hence, for the whole range of scales where we know the observations give a non-zero primordial 4PCF, our asymptotic is good and the odd CMB trispectrum corresponding to such an odd LSS 4PCF would be suppressed. We note that this argument still relies on our being in the low-$\ell$, SW limit, as we needed the sBfs from the SW-limit transfer functions to make this argument.

\bibliography{ref}

\end{document}